\documentclass[preprint,12pt]{elsarticle}

\usepackage{graphicx}
\usepackage{amssymb}
\usepackage{lineno}
\usepackage[bookmarks,bookmarksnumbered]{hyperref}
\hypersetup{colorlinks = true,linkcolor = blue,anchorcolor =red,citecolor = blue,filecolor = red,urlcolor = red}
\usepackage{xcolor}
\usepackage{graphicx}
\usepackage[version=4]{mhchem}
\usepackage{amssymb}
\usepackage{amsmath}
\def\eps{\ensuremath\varepsilon}

\usepackage{lineno}
\usepackage[margin=1.0in]{geometry}
\usepackage{natbib}
\usepackage{enumitem}
\newtheorem{remark}{Remark}

\journal{arXiv}

\begin{document}

\begin{frontmatter}

%% Title, authors and addresses

%\title{The slow scale linear noise approximation \` a la Fenichel theory: 
%Stochastic enzyme kinetics as a case study}

\title{Stochastic enzyme kinetics and the quasi-steady-state reductions: 
Application of the slow scale linear noise approximation \` a la Fenichel}

%% use the tnoteref command within \title for footnotes;
%% use the tnotetext command for the associated footnote;
%% use the fnref command within \author or \address for footnotes;
%% use the fntext command for the associated footnote;
%% use the corref command within \author for corresponding author footnotes;
%% use the cortext command for the associated footnote;
%% use the ead command for the email address,
%% and the form \ead[url] for the home page:
%%
%% \title{Title\tnoteref{label1}}
%% \tnotetext[label1]{}
%% \author{Name\corref{cor1}\fnref{label2}}
%% \ead{email address}
%% \ead[url]{home page}
%% \fntext[label2]{}
%% \cortext[cor1]{}
%% \address{Address\fnref{label3}}
%% \fntext[label3]{}

%% use optional labels to link authors explicitly to addresses:
%% \author[label1,label2]{<author name>}
%% \address[label1]{<address>}
%% \address[label2]{<address>}

\author[label1]{Justin Eilertsen}
\author[label2]{Kashvi Srivastava}
\author[label3,label4]{Santiago Schnell}
\ead{santiago.schnell@nd.edu}
\address[label1]{Mathematical Reviews, American Mathematical Society, 416 $4th$ Street, 
Ann Arbor, MI 48103, USA}
\address[label2]{Department of Mathematics, University of Michigan, Ann Arbor, MI 48109, 
USA}
\address[label3]{Department of Biological Sciences, University of Notre Dame, Notre Dame, 
IN 46556, USA}
\address[label4]{Department of Applied and Computational Mathematics and Statistics, 
University of Notre Dame, Notre Dame, IN 46556, USA}

\begin{abstract}
The linear noise approximation models the random fluctuations from the mean-field model 
of a chemical reaction that unfolds near the thermodynamic limit. Specifically, the 
fluctuations obey a linear Langevin equation up to order $\Omega^{-1/2}$, where $\Omega$ 
is the size of the chemical system (usually the volume). In the presence of disparate
timescales, the linear noise approximation admits a quasi-steady-state reduction referred 
to as the \textit{slow scale} linear noise approximation (ssLNA). Curiously, the ssLNAs 
reported in the literature are slightly different. The differences in the reported ssLNAs 
lie at the mathematical heart of the derivation. In this work, we derive the ssLNA directly 
from geometric singular perturbation theory and explain the origin of the different ssLNAs 
in the literature. Moreover, we discuss the loss of normal hyperbolicity and we extend 
the ssLNA derived from geometric singular perturbation theory to a non-classical singularly 
perturbed problem. In so doing, we disprove a commonly-accepted qualifier for the validity 
of stochastic quasi-steady-state approximation of the Michaelis--Menten reaction mechanism. 
\end{abstract}


\end{frontmatter}

%%
%% Start line numbering here if you want
%%
%% \linenumbers

%%%%%%%%%%
\section{Introduction} \label{S:into}
The set of elementary reactions that comprise a chemical system often occur at disproportionate 
rates. From the chemical physics point of view, chemical systems whose 
elementary reaction rates are disparate constitute a \textit{multiscale} process. From 
a modeling point of view, multiscale reactions are highly advantageous, since the 
presence of widely separated timescales permits a reduction in the number of mathematical 
equations required to model the reaction over slow (long) timescales.

In the deterministic regime, near the thermodynamic limit, chemical equations can be 
accurately modeled with a system of nonlinear ordinary differential equations. 
The reduction of deterministic models is generally achieved through the application 
of Tikhonov's theorem \cite{Tikhonov1952} and Fenichel theory~\cite{Fenichel1971,Fenichel1979}. 
Several analyses of enzyme-catalyzed reactions have made good use of singular 
perturbation theory to generate approximations referred to as \textit{quasi-steady-state} 
(QSS) approximations or reductions \cite{Eilertsen2019,EILERTSEN2020}. In fact, over the 
last decade, much progress has been made in developing and applying the formalism of Fenichel  
theory to chemical kinetics, and the culmination of the recent literature has turned 
up some surprising results. First and foremost, the advent of Tikhonov-Fenichel 
parameter value (TFPV) theory, developed extensively by 
\citet{Goeke2017,Goeke2015}, has rigorously demonstrated that not 
all QSS reductions emerge as a result of a singular perturbation scenario, 
despite what scaling and numerical simulations might suggest~\cite{Noethen2011,OpenMMin}. 
TFPV theory has also enhanced our understanding of the singular perturbation 
structure (when applicable) to pertinent reaction models, which has led to 
the discovery of bifurcations and other interesting phenomena present in the 
singular vector fields of the model equations~\cite{EILERTSEN2020}. Most 
surprising, however, is the revelation that traditional scaling methods may 
lead to erroneous conclusions concerning the mathematical origin and 
justification of QSS reduction in chemical kinetics (see, for example, 
Goeke et al.~\cite{Goeke2012}, Section 4, as well as \citet{OpenMMin}).

Given the recent developments in the deterministic theory of QSS reduction, the natural 
question to ask is: \textit{Do any of these developments have something 
important to say about model reduction in the stochastic realm?} Model reduction 
is more challenging in the stochastic regime, but rigorous reduction methods that 
leverage the presence of disparate timescales do exist (see, for 
example, \cite{kan,KangKim2017,kang2013}). The focus of this paper is on the application 
of QSS reduction in the linear noise regime, where stochastic fluctuations from the 
deterministic mean-field model are governed by a linear Langevin equation called 
the \textit{linear noise approximation} (LNA). The general methodology for 
QSS reduction in the LNA regime, called the \textit{slow scale linear noise 
approximation} (ssLNA), is by \citet{Thomas2012},
\citet{PAHLAJANI201196}, and \citet{Herath}. Interestingly, the reported ssLNAs 
are slightly different, and this raises the question: Where do these differences 
come from, and are they critical? 
The intent of this paper is three-fold: (i) to explain why different ssLNAs 
exist in the literature, (ii) to introduce recent developments of deterministic 
QSS theory to the stochastic community, and (iii) to demonstrate techniques to 
extend the ssLNA to specific non-classical singularly-perturbed problems. In what 
follows, we revisit the mathematical formalism of geometric singular perturbation 
theory (GSPT) and derive the ssLNA directly from GSPT. We discuss the role of TFPV 
theory in the applicability of GSPT, and demonstrate where the differences emerge 
between the GSPT-derived ssLNA and the ssLNAs of \citet{Thomas2012}, 
\citet{PAHLAJANI201196}, and \citet{Herath}. We also 
discuss the role of the GSPT-derived ssLNA in the QSS reduction of 
the chemical master equation (CME) and use it to debunk a well-established result 
in the literature. 

%%%%%%%%%%
\section{Singular perturbations and Fenichel Theory: A brief introduction}\label{sec1}
In this section, we give a very brief overview of Fenichel theory as it applies to 
singular perturbations by shadowing \citet[][Chapter 3]{Wechselberger2020}. 
However, the results were originally obtained by \citet[][Section 5]{Fenichel1979}. 
A detailed mathematical expose on Fenichel reduction and its applicability 
in enzyme kinetics can be found in \cite{Noethen2011, Goeke2012, Goeke2014}.

%%%%%
\subsection{Coordinate-free slow manifold projection}
Fenichel theory is concerned with the persistence of normally hyperbolic invariant 
manifolds with respect to a perturbation. Dynamical systems subject to a small 
perturbation are of the general form
\begin{equation}\label{dy}
    \dot{z}= w(z) + \eps G(z,\eps)
\end{equation}
 where $0<\eps \ll 1$. The stationary points of the unperturbed vector field, $w(z)$, 
 determine the classification of the perturbation problem. If the perturbation is 
 singular, then there exists a set, $S$, comprised of non-isolated equilibrium points: 
 \begin{equation*}
 S:=\{z\in \mathbb{R}^n: w(z)=0\}.
 \end{equation*}

Fenichel reduction applies to compact subsets, $\mathcal{S}_0 \subseteq S$ that are 
differentiable manifolds (with a possible boundary). The compactness requirement of 
$\mathcal{S}_0$ is generally easy to satisfy in chemical kinetics: due to conservation 
laws, phase-space trajectories remain within a bounded, positively invariant set, 
$\Lambda$. If $S_0$ is an embedded $k$-dimensional submanifold of $\mathbb{R}^n$, then
\begin{equation*}
    \text{rank}\;Dw(z) = n-k \quad \forall z\in S_0.
\end{equation*}
Furthermore, if the algebraic and geometric multiplicity of the zero eigenvalue are 
both equal to $k$, then
\begin{equation}
    T_zS_0 :=\{x\in \mathbb{R}^n: Dw(z)\cdot x = 0\} = \ker Dw(z) \quad \forall z \in S_0,
\end{equation}
and there is
continuous splitting,
\begin{equation}\label{splitting}
   \mathbb{R}^n := \ker Dw(z) \oplus \text{Image}\;Dw(z)
\end{equation}
for all $z \in S_0$. Perturbing the vector field by setting $0 < \eps \ll 1$ results 
in the formation of an invariant, slow manifold, $M$.  If the real parts of the $n-k$ 
non-zero eigenvalues of $Dw(z)$ are strictly less than zero,\footnote{We will assume 
this to hold throughout so that both the critical and slow manifolds are attracting.} 
then $M$ will attract nearby trajectories at an exponentially fast rate. Projecting 
the perturbation, $\varepsilon G(z,0)$, onto the tangent space of $S_0$ results in a 
reduced equation (called a QSS approximation) that captures the 
long-time behavior of the system.  

\begin{figure}[htb!]
    \centering
    \includegraphics[scale=2.5]{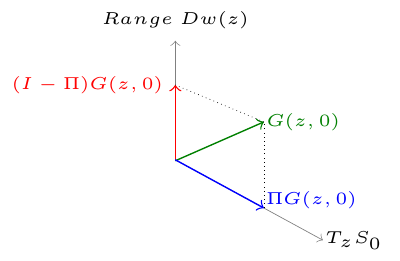}
    \caption{\textbf{The geometry of $\Pi^{S_0}$.} 
    \label{fig:oblique} The projection matrix, labeled here as $\Pi$, defines an oblique projection: 
    while ${\rm Range} \; Dw(z) \cap \ker Dw(z) = \{0\}$ for $z\in S_0$, ${\rm Range} \;Dw(z)$ is not 
    necessarily     orthogonal to $\ker Dw(z)$.}
\end{figure}

The decomposition (\ref{splitting}) implies the existence of a projection operator, 
$\Pi^{S_0}$, that maps to $\ker Dw(z)$
\begin{equation}
    \Pi^{S_0}:\mathbb{R}^n \mapsto T_zS_0 \quad \forall x \in S_0.
\end{equation}
The explicit form of $\Pi^{S_0}$ is obtained 
by exploiting the fact that $w(z)$ factors (locally) as
\begin{equation}\label{feq}
   w(z):=N(z)\mu(z), \quad N(z)\in \mathbb{R}^{n \times (n-k)}, \quad \mu(z)\in \mathbb{R}^{n-k}.
\end{equation}
The columns of $N$ comprise a basis for the range of the Jacobian, $Dw(z)$, and the zero level 
set of $\mu(z)$ is identically $S_0$. Since $\text{rank} \;Dw(z)=n-k$, and the 
zero set of $\mu:\mathbb{R}^n\mapsto \mathbb{R}^{n-k}$ corresponds to $S_0$ (a submanifold of 
$\mathbb{R}^n$), we have that, for all $z\in S_0$, $N(z)$ has full (column) rank, and $D\mu(z)$ has full (row) rank:
\begin{subequations}
\begin{align}
\text{rank}\; &N(z) = \;n-k,\\
    \text{rank}\; &D\mu(z) = n-k.
    \end{align}
\end{subequations}
The row vectors of $D\mu(z)$ form a basis 
for the orthogonal complement of $\ker Dw(z)$. Since projection operators are uniquely determined 
by their range and the orthogonal complement of their kernel,
the operator $\Pi^{S_0}$ is
\begin{equation}
 \Pi^{S_0} := I- N(D\mu N)^{-1}D\mu
\end{equation}
which is an \textit{oblique} 
projection operator (see {{\sc Figure}}~\ref{fig:oblique} for a geometric interpretation of $\Pi^{S_0}$).
%%%%%

Once the projection operator is constructed, the reduced equation is formulated by projecting the 
perturbation, $G(z,0)$, onto $\ker Dw(z)$:
\begin{equation*}
    \dot{z} = \Pi^{S_0}G(z,0)|_{z\in S_0}.
\end{equation*}

%%%%%%%%%%
\section{Singular perturbation reduction in biochemical kinetics: Didactic 
examples}\label{sec2}
In this section, we compute several QSS reductions of the MM reaction mechanism. We introduce the mass 
action equations of the deterministic MM reaction mechanism and discuss the computation of QSS reductions 
directly from Fenichel theory without a~priori non-dimensionalization. Several QSS reductions are computed, 
including the standard QSS approximation (sQSSA) and the quasi-equilibrium approximation (QEA).

%%%%
\subsection {The Michaelis--Menten reaction mechanism}
The MM reaction consists of three elementary reactions: the binding of a  substrate molecule, S, with 
an enzyme molecule, E, leading to the formation of an intermediate complex molecule, C. The complex 
molecule can disassociate back into unbound enzyme and substrate molecules, or it disassociates into 
a product molecule, P, and an enzyme molecule. The chemical  equation is given by
\begin{equation}\label{mm1}
    \ce{S + E <=>[$k_1$][$k_{-1}$] C ->[$k_2$] E + P },
\end{equation}
where $k_1$, $k_{-1}$ and $k_2$ are deterministic rate constants.

The mass action equations that describe the kinetics of (\ref{mm1}) in the 
thermodynamic limit constitute a two-dimensional system of nonlinear ordinary
differential equations,
\begin{subequations}\label{mmMASS}
\begin{align}
    \dot{s} &= -k_1(e_T-c)s+k_{-1}c,\label{m1}\\
    \dot{c} &= k_1(e_T-c)s -(k_{-1}+k_2)c\label{m2},
\end{align}
\end{subequations}
where lowercase $s$, $c$, $e$ and $p$ denote the concentrations of S, C,
E and P, respectively. Once the temporal dynamics of $s$ and $c$ are known, 
the evolution of product is recovered from
\begin{equation}
    \dot{p}=k_2c\label{m3}.
\end{equation}
The temporal concentration of enzyme, $e$, is computed from $e_T-c$, where $e_T$ 
is a conserved quantity, the \textit{total} enzyme concentration, and accounts for
the concentration of both bound and unbound enzyme molecules. A second conservation 
law is obtained from the addition of (\ref{m1})--(\ref{m3}), $\dot{s}+\dot{c}+\dot{p}=0$, 
yielding the conservation of substrate:
\begin{equation}
    s_T = s+c+p.
\end{equation}
Unless otherwise stated, we will take $s(0)=s_T$ in the analysis that follows, 
which implies $c(0)=p(0)=0$. 

%%%%%
\subsection{Tikhonov-Fenichel Parameter Value Theory}
It is possible (and convenient) to compute QSS reductions directly from the dimensional equation. 
This a result of the TFPV theory developed by Goeke et al.~\cite{Goeke2017,GoekeDis, GOEKEpara}, 
which we briefly outline here. 
 
In physical applications, most dynamical systems depend on an $m$-tuple of parameters, 
$\pi \in \mathbb{R}^m$:
\begin{equation*}
 \dot{z}=f(z,\pi), \quad z\in \mathbb{R}^n, \;\;\pi\in \mathbb{R}^m, \;\; f:\mathbb{R}^n\times \mathbb{R}^m \mapsto \mathbb{R}^n.
\end{equation*}
A TFPV value is a point, $\widehat{\pi}$, in parameter space for which the vector field, 
$f(z,\widehat{\pi})$, contains a normally hyperbolic and attracting critical manifold. 
 
As an example, the MM reaction mechanism mass action equations depend on the parameters 
$\pi =(e_T,k_1,k_{-1},k_2)^{tr.}$, where $tr.$ denotes transpose. There are three engaging 
TFPV values\footnote{The non-zero parameters in $\widehat{\pi}$ are appropriately bounded 
below and above by positive constants.} associated with the MM reaction mechanism:
\begin{subequations}
\begin{align*}
    \widehat{\pi}_1:=(0,k_1,k_{-1},k_2)^{tr.},\\
    \widehat{\pi}_2:=(e_T,0,k_{-1},k_2)^{tr.},\\
    \widehat{\pi}_3:=(e_T,k_1,k_{-1},0)^{tr.}.
\end{align*}
\end{subequations}
%%%%%%%%%%
Singular perturbation theory applies to vector fields that are sufficiently close to the 
TFPVs. Thus, the QSS reductions that are constructed by projecting onto the tangent space 
of a critical manifold associated with the TFPVs will be valid for $\pi$ sufficiently 
close $\widehat{\pi}_i$. Consequently, we will consider parameter values close to TFPVs:
\begin{subequations}
\begin{align*}
    {\pi}_1:=(\eps \widehat{e}_T,k_1,k_{-1},k_2)^{tr.},\\
    {\pi}_2:=(e_T,\eps \widehat{k}_1,k_{-1},k_2)^{tr.},\\
    {\pi}_3:=(e_T,k_1,k_{-1},\eps \widehat{k}_2)^{tr.},
\end{align*}
\end{subequations}
where $\eps$ is very small but positive, and $\widehat{e}_T$, $\widehat{k}_1$ and $\widehat{k}_2$ 
are of unit magnitude and simply encode the units of $e_T$, $k_1$ and $k_2$, respectively. As we 
demonstrate in the subsection that follows, this notation enables the computation of QSS 
reductions without the need to non-dimensionalize the mass action equations.

%%%%%
\subsection{Fenichel reduction: The sQSSA of the MM reaction mechanism}
To extract a reduced model from (\ref{mmMASS}), we begin with the assumption that $e_T$ 
is small and therefore $\pi$ is close to $\widehat{\pi}_1$. Consequently, we rescale $e_T$ 
as $e_T \mapsto \eps \widehat{e}_T$, where $0<\eps \ll 1$ (again, this notation really just 
serves as a reminder that $e_T$ is \textit{small}). In $(s,c)$ coordinates, we have 
$z:=(s\;\;c)^T$, and in perturbation form, the mass action equations~(\ref{mmMASS}) are
\begin{equation}\label{p1}
    \dot{z} = w(z) + \eps G(z,\eps), \quad w(z):=\begin{pmatrix}k_1cs +k_{-1}c\\ -k_1cs -(k_{-1}+k_2)c\end{pmatrix}, \quad G(z,\eps):= \begin{pmatrix}-k_1\widehat{e}_Ts\\ \;\;k_1\widehat{e}_Ts\end{pmatrix}.
\end{equation}
The 
singular problem recovered by setting $\eps=0$ so that $\pi=\widehat{\pi}_1$ yields a 
critical manifold of equilibria
\begin{equation}
    S_0:=\{(s,c)\in\mathbb{R}^2: c=0, \quad 0 \leq s \leq s_T\}.
\end{equation}
It is straightforward to verify that $S_0$ is normally hyperbolic. Moreover, since the 
non-trivial eigenvalue of the Jacobian, $\lambda_{MM}$, is strictly less than zero
\begin{equation*}
    \lambda_{MM} :=-(k_1s+k_{-1}+k_2)
\end{equation*}
the critical manifold is attractive. 

Since $S_0$ is normally hyperbolic and attracting, we proceed to compute $\Pi^{S_0}$. The 
factorization of $w(z)$ is straightforward to compute
\begin{equation}\label{23}
    w(z)=N(s,c)\mu(s,c), \quad \text{with}\quad N(s,c):= \begin{pmatrix}k_1s + k_{-1}\\ -k_1s -k_{-1}-k_2\end{pmatrix}, \quad\text{and} \quad \mu(s,c):=c,
\end{equation}
as is the derivative of $\mu(s,c)=c$:
\begin{equation}
    D\mu(s,c) = (0\;\; 1).
\end{equation}
Putting the pieces together, the projection operator $\Pi^{S_0}$ is 
\begin{equation}\label{proj1}
    \Pi^{S_0}:=\begin{pmatrix}1 & \gamma(s) \\ 0 & 0\end{pmatrix}, \quad \gamma(s):=\cfrac{K_S+s}{K_M+s},
\end{equation}
where $K_S = k_{-1}/k_1$ and $K_M = (k_{-1} + k_2)/k_1$.
The corresponding QSS approximation is
\begin{equation}\label{sQSSA}
\dot{s}=\Pi^{S_0}G(s,c,0)|_{c=0}:=-\cfrac{k_2e_Ts}{K_M+s},
\end{equation}
which is the sQSSA.\footnote{In (\ref{sQSSA}), 
we have transformed $\eps \widehat{e}_T$ back to $e_T$ for clarity, and will continue 
to do this from this point forward.}

\begin{remark}
Note that the sQSSA (\ref{sQSSA}) is not the result of singular perturbation problem 
that is in standard form
\begin{subequations}
\begin{align*}
    \dot{x} &= \eps f(x,y,\eps),\\
    \dot{y} &= \;\;g(x,y,\eps).
\end{align*}
\end{subequations}
This is contrary to the justification established from scaling analyses that utilize 
non-dimensionalization (see, \citet{Heineken1967,Segel1989}).
\end{remark}

%%%%%
\subsection{Fenichel reduction: The QEA}
In addition to the sQSSA, the QEA is a QSS reduction that is valid in the limit of slow product 
formation that occurs when $\pi$ is close to $\widehat{\pi}_3$. Rescaling $k_2$ as 
$k_2 \mapsto \eps \widehat{k}_2$, the mass action system
\begin{subequations}
\begin{align}
\dot{s} &= -k_1(e_T-c)s+k_{-1}c,\\
\dot{c} &= k_1(e_T-c)s -k_{-1}c - \eps \widehat{k}_2 c,
\end{align}
\end{subequations}
has a critical manifold of equilibria in the singular limit that coincides with $\pi =\widehat{\pi}_3:$ 
\begin{equation}\label{critk2}
    S:=\bigg \{(s,c)\in \mathbb{R}^2: c= \cfrac{e_Ts}{K_S+s} \bigg\},
\end{equation}
which is identical to the $s$-nullcline. The QEA in $(s,c)$ coordinates is well understood but 
trickier than the sQSSA. The consequence is that there can be noticeable depletion of $s$ during 
the approach to the slow manifold unless $e_T \ll K_M + s_T$~\cite{Segel1988,Segel1989}. We will 
not rehash the details here, but state the main results also found in \cite{Goeke2017,Wechselberger2020}. 
The projection matrix, $\Pi^{S_0}$, and perturbation, $G(s,c,0)$, are given by
\begin{equation}\label{sp}
    \Pi^{S_0}:=\cfrac{1}{(e_T-c+K_S+s)}\begin{pmatrix}(K_S+s) & (K_S+s)\\ (e_T-c) & (e_T-c)\end{pmatrix}, \qquad G(s,c,0):=-\begin{pmatrix}0\\\widehat{k}_2c\end{pmatrix},
\end{equation}
and corresponding QSS reduction for $s$ is
\begin{equation}\label{slowp}
\dot{s}=-\cfrac{k_2e_Ts(K_S+s)}{e_TK_S+(K_S+s)^2}, \quad \dot{p}=\cfrac{k_1k_2e_Ts}{k_1s+k_{-1}}.
\end{equation}

%%%%%
\subsection{Fenichel reduction: The reverse QSSA}
The reverse QSSA (rQSSA) was originally defined by \citet{Segel1989} as a perturbation problem, and 
later investigated in detail by \citet{SCHNELL2000}.  To preface the derivation of the rQSSA 
as a Fenichel reduction, we remark that there are two common conditions 
that emerge in biochemical applications:
\begin{enumerate}
    \item $ND\mu(z)$ vanishes (or changes rank) at at least one point belonging to the critical set. This happens, for example, if $D\mu(z) =0$ at some point belonging to the set $\mu(z)=0$.\
    \item The zero eigenvalue of the Jacobian evaluated at at least one point in $S_0$ has an algebraic 
    multiplicity that is greater than the geometric multiplicity (the splitting (\ref{splitting}) does 
    not hold at such points).
\end{enumerate}

\begin{figure}[htb!]
    \centering
    \includegraphics[scale=1.75]{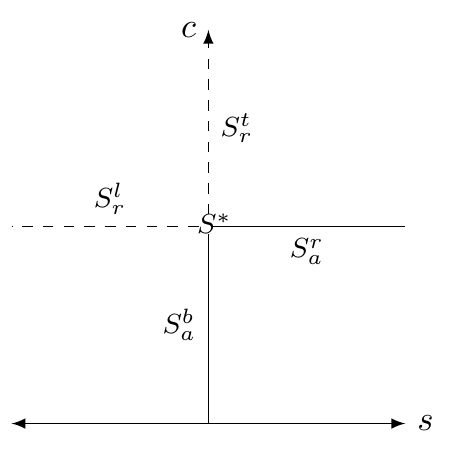}
    \caption{\textbf{A dynamic transcritical bifurcation occurs in the singular 
    limit corresponding to $k_2=k_{-1}=0$ of the MM  reaction
    mechanism.}  At $S^{\star}$ the Jacobian has a double-zero eigenvalue. Along the dashed lines the 
    Jacobian has one zero eigenvalue and one positive eigenvalue. Along the solid lines the Jacobian 
    has one zero eigenvalue and one negative eigenvalue. At the bifurcation point $S^{\star}=(0,e_T)$, 
    the lines     $S_a^r \cup S_r^l$ and $S_r^t \cup S_a^b$ intersect and exchange stability.
    \label{fig:CRTMNF}}
\end{figure}

The rQSSA is valid for small $k_{-1}$ \textit{and} small $k_2$, and is of the variety 1. In perturbation 
form this corresponds to
\begin{subequations}
\begin{align}
    \dot{s} &= -k_1(e_T-c)s+\varepsilon \widehat{k}_{-1}c,\\
    \dot{c} &= k_1(e_T-c)s-\varepsilon(\widehat{k}_{-1}+\widehat{k}_2)c.
\end{align}
\end{subequations}
The critical set is given by,
\begin{equation}
    S_0 :=\{(s,c)\in \mathbb{R}^2_{\geq 0}: c=e_T, 0 \leq s \leq s_T\}\cup \{(s,c)\in \mathbb{R}^2_{\geq 0}: s=0, 0 \leq c \leq e_T\}.
\end{equation}
The rank of the Jacobian along $S_0$ is not constant
\begin{subequations}
\begin{align*}
{\rm rank}\;\; Dw(s,c) &=0,\;\;\text{if}\;\;(s,c)=(0,e_T),\\
{\rm rank}\;\; Dw(s,c)&= 1, \;\;\text{otherwise},
\end{align*}
\end{subequations}
and thus TFPV theory does not apply.\footnote{The point $\pi^*=(e_T,k_1,0,0)^{tr.}$ is not a TFPV.} 
However, observe that the compact submanifolds
\begin{subequations}
\begin{align*}
S_a^r&:=\{(s,c)\in \mathbb{R}^2_{\geq 0}: c=e_T,\;\; \varrho \leq s \leq s_T\}, \qquad\; 0 < \varrho,\\
S_a^b&:=\{(s,c)\in \mathbb{R}^2_{\geq 0}: s=0,\;\; 0 \leq c \leq e_T-\kappa\}, \quad 0 < \kappa < e_T
\end{align*}
\end{subequations}
are normally hyperbolic and attracting. When $s(0)>0$, trajectories will initially approach and follow 
$S_a^r$ before eventually following $S_a^b$. In fact, a quick analysis reveals the existence 
of a transcritical bifurcation (see {\sc Figure}~\ref{fig:CRTMNF}).
%%%%%

By the projection methods above, it is straightforward to show that the QSS reductions obtained via projection 
onto $T_zS_a^r$ and $T_zS_a^b$ are, respectively:
\begin{subequations}
\begin{align}
    \dot{p} &= k_2e_T, \qquad \qquad \;\;  0 \leq p < s_T-e_T\\
    \dot{p} &= k_2(s_T-p), \qquad s_T-e_T < p \leq s_T.
\end{align}
\end{subequations}

As a concluding remark, note that we have successfully computed QSS reductions without a priori scaling 
and non-dimensionalization of the mass action equations. The ability to compute QSS reductions directly 
from the dimensional equations is a result of the TFPV theory developed by \citet{Goeke2015,Goeke2017}, 
which we have utilized here.

%%%%%%%%%%
\section{Stochastic chemical kinetics: Expansions, reductions, and approximations}\label{sec4}
In this section, we discuss QSS reduction in the stochastic regime. We introduce the CME and 
the derivation of the LNA via the $\Omega$--expansion. We conclude with 
a review of the ssLNA as derived by \citet{Thomas2012} and \citet{PAHLAJANI201196}, 
and we compare it to the GSPT-derived ssLNA. 

%%%%%
\subsection{Stochastic chemical kinetics far from the thermodynamic limit: The master equation}
Under physical conditions, a reaction occurs within a bounded volume, $\Omega$. 
If the number of molecules in the system is finite, the reaction will \textit{always} 
exhibit fluctuations. In fact, in the presence of random fluctuations and intrinsic 
noise, stochastic models provide a more physically realistic description of the kinetics 
when a system is far from the thermodynamic limit since the time interval between 
successive reactions becomes a random variable. The appropriate mathematical model depends on 
how ``close" the system is to the thermodynamic limit. 

If the chemical reaction consists of ``$R$" elementary reactions, and the mixture is 
homogeneous and not diffusion limited then, far from the thermodynamic limit, the probability 
of finding the system in state ${Z}$ at time $t$ can be obtained from the solution to 
CME (see, \cite{GILLESPIECME} and \cite{VKmaster} for details),
\begin{equation}\label{CME}
    \cfrac{\partial P(Z,t)}{\partial t} = \sum_{j=1}^Ra_j(Z-{\nu}_j)P({Z}-{\nu}_j,t)-a_j({Z})P({Z},t),
\end{equation}
where ${\nu}_j$ are the stoichiometric vectors that correspond to the $jth$ elementary 
reaction. If the system is in state $Z$ when the $jth$ reaction occurs, then the new state 
of the system will be ${Z} + {\nu}_j$. The functions $a_j$ are called 
\textit{propensity functions}. Dynamically the state of the system at time $t$ is ${Z}$,
and it moves from state ${Z}$ to the state ${Z}+{\nu}_j$ within the infinitesimal 
window $[t,t+\text{d}t)$ with the probability
 \begin{equation}\label{cond}
     P({Z}+{\nu}_j,t+\text{d}t |{Z},t) = a_j({Z})\text{d}t.
 \end{equation}
The conditional probability~(\ref{cond}) of jumping into the state ${Z}+{\nu}_j$ depends 
only on the present state of the system, which is called the Markov property.

The CME is possibly the most fundamental description of a chemical reaction. The difficulty is 
that closed-form solutions are rarely attainable. This begs question: Is it possible to derive 
physical models that exhibit stochasticity, but are nevertheless easier to analyze? The answer 
is yes, but the cost is that simplified models are usually only valid in monostable systems near 
the thermodynamic limit. The LNA is of this variety. 

%%%%%
\subsection{Approaching the thermodynamic limit: the LNA}
To introduce the LNA, it is helpful to express the mass action equations in the form
\begin{equation}\label{comMASS}
    \dot{z} = \mathcal{S}q,\quad F:=diag(q)
\end{equation}
where $\mathcal{S}$ is the stoichiometric matrix, and $q$ is the main diagonal of the 
matrix $F$, whose diagonal components correspond to the elementary reactions of the 
chemical system. For example, the MM reaction mechanism~(\ref{mm1}) consists of three 
elementary reactions: the formation of complex, the disassociation of complex into S 
and E, and the disassociation of complex into E and P. Hence, the mass action 
system in 
form (\ref{comMASS}) is
\begin{equation}\label{MMfull}
    \begin{pmatrix}\dot{s}\\\dot{c}\end{pmatrix} = \begin{pmatrix}-1 & \;\;\;1 & \;\;\;0\\\;\;\;1 & -1 & -1\end{pmatrix}\begin{pmatrix}k_1(e_T-c)s\\ k_{-1}c\\k_2c\end{pmatrix}:=\mathcal{S}q.
\end{equation}
To formally derive the LNA, one starts with the \textit{operator form} of the CME,
\begin{equation}\label{sCME}
    \cfrac{\partial P({Z},t)}{\partial t} =\Omega \sum_{j=1}^R\bigg(\prod_{i=1}^m \mathbb{E}^{\mathcal{S}_{ij}}-1\bigg)a_j({z})P({Z},t),
\end{equation}
where $z=Z/\Omega$ and $\mathbb{E}^{\mathcal{S}_{ij}}$ is the step 
operator:\footnote{Here, $e_i$ is the standard basis vector in $\mathbb{R}^n.$}
\begin{equation}\
    \mathbb{E}^{-\mathcal{S}_{ij}}a({Z}) = a(\mathbb{E}^{-\mathcal{S}_{ij}} {Z})= a({Z}-\mathcal{S}_{ij}{e}_i).
\end{equation}
Inserting the ansatz ${Z} = \Omega {z} + \Omega^{1/2}{X}$ into (\ref{sCME}) and 
expanding (\ref{sCME}) in powers of $\Omega$ yields (\ref{comMASS}) at zeroth-order 
in $\Omega$. Thus, the mean of the stochastic trajectory obeys the mass action 
equation~(\ref{comMASS}). 

At order $\Omega^{-1/2}$, the equation that determines the randomly fluctuating departure 
from the mean, $X$, is a linear SDE,
\begin{equation}\label{SDE1}
\text{d}{X} =J{X}\;\text{d}t + \Omega^{-1/2}\mathcal{S}\sqrt{F}\;\text{d}{W},
\end{equation}
where $J$, the Jacobian, is $J:=D(Sq)$, and $W$ is a Wiener process. Collectively, (\ref{SDE1}) and 
the mass action equations comprise the LNA. On occasion we will express the LNA in the form
\begin{equation}\label{SDEZ}
\dot{X} =J{X}+ \Omega^{-1/2}\mathcal{S}\sqrt{F}\;\zeta(t),
\end{equation}
where the Gaussian white noise, $\zeta(t)$,
is understood to be the generalized derivative of $W$.

The LNA is notably simpler than the CME, 
since the Langevin equation (\ref{SDE1}) is \textit{linear}, and the integration of 
linear stochastic differential equations of the form~(\ref{SDE1}) is well-understood. 
The Fokker-Plank equation associated with (\ref{SDE1}) is 
also linear,
\begin{equation}\label{FP1}
\cfrac{\partial \rho({X},t)}{\partial t} =\bigg( -\cfrac{\partial }{\partial X_i}(J {X})_i + \cfrac{1}{2\Omega} D_{ij} \cfrac{\partial}{\partial X_i}\cfrac{\partial}{\partial X_j}   \bigg) \rho({X},t),
\end{equation}
where the diffusion matrix, $D$, is given by $ D = \mathcal{S} F \mathcal{S}^T$.

As mentioned in the earlier sections, the reduction of the LNA based on timescale separation 
is the ssLNA developed by \citet{Thomas2012,ThomasPO} and \citet{PAHLAJANI201196}. In the nonlinear 
regime, \citet{Katz} addressed reduction of stochastic differential equations (SDEs) 
of the form
\begin{equation}\label{katzEQ}
    \text{d}{x} = w(x)\text{d}t + \varepsilon G(x,\varepsilon)\text{d}t + \sqrt{\nu} B(x)\;\text{d}W
\end{equation}
where $\varepsilon$ and $\nu$ are extremely small (i.e., $\eps, \nu \ll 1$). In short, \citet{Katz} 
proved that provided specific conditions hold, SDEs of 
the form (\ref{katzEQ}) converge, in a certain sense, to the reduced SDE,
\begin{equation}
    \text{d}x = \varepsilon \Pi G(x,\varepsilon)\text{d}t +  \sqrt{\nu} \Pi B(x)\;\text{d}W + \nu \mathfrak{D}(x,\eps),
\end{equation}
where $\Pi$ is a projection operator that maps to the tangent space of the critical manifold 
$S_0:=\{x\in \mathbb{R}^n:w(x)=0\}$ that emerges when $\eps = \nu =0$, and $\nu \mathfrak{D}(x)$ 
is a noise-induced drift term. \citet{Parsons2017} derived the explicit construction of 
$\Pi$ and $\mathfrak{D}(x)$  in their analysis of fully nonlinear Langevin equations. Notably, 
\citet{Parsons2017} did not discuss the reduction of noisy systems in standard form, and a 
projection operator $\Pi$ that is consistent with Fenichel theory has not been defined for 
standard-form singularly perturbed systems in the linear noise regime. Such is the subject 
of the subsection that follows.

%%%%%
\subsection{Projecting onto the tangent space of the critical manifold}
Classical singular perturbation reduction of a deterministic system requires the existence of a normally 
hyperbolic critical manifold in the singular limit; this fact is non-negotiable. The reduction 
of the LNA is also straightforward, provided one has a well-defined critical manifold. The key 
observation in the LNA regime is to recognize that the dimension of the problem increases, but 
that the LNA is still of the form (\ref{katzEQ}), and therefore the results of \citet{Katz} 
are applicable. All that remains is to identify a normally hyperbolic critical manifold, its 
tangent space, and construct the unique projection operator, $\Pi$. 

In the standard form, the general LNA is\footnote{The Jacobian of the layer problem is equal to $ND\mu$ whenever $z\in S_0$.}
\begin{subequations}\label{SSlna}
\begin{align}
\dot{z} &= w(z) + \varepsilon G(z,\varepsilon),\\
\dot{X} &=ND\mu\cdot {X} + \varepsilon DG(z,\varepsilon)\cdot X\; + \Omega^{-1/2}\mathcal{S}\sqrt{F}\cdot\Gamma, 
\end{align}
\end{subequations}
where $\Gamma:= (\zeta_1(t), \zeta_2(t),..)^T$ is a white noise vector:
\begin{equation*}
    \langle \zeta_i(t),\zeta_j(\tau)\rangle = \delta_{ij}(t-\tau).
\end{equation*}
In \textit{perturbation form}, the LNA is
\begin{equation}\label{GLNA}
    \dot x = \widetilde{w}(x) + \varepsilon \boldsymbol{\tilde{G}}(x,\varepsilon) + \Omega^{-1/2}B(z)\Gamma, 
\end{equation}
with
\begin{equation}
x := \begin{pmatrix} z \\ X \end{pmatrix}, \quad \widetilde{w}(x):=\begin{pmatrix} N\mu \\ND\mu \cdot{X}
\end{pmatrix}, \quad \boldsymbol{\tilde{G}}(x,\varepsilon) := \begin{pmatrix} G(z,\varepsilon)\\DG(z,\varepsilon)\cdot X
\end{pmatrix}, \quad B:=\begin{pmatrix}\mathbb{O}^{n\times m} \\ S\sqrt{F}\end{pmatrix}.
\end{equation}

For a planar system in which $z\in \mathbb{R}^2$, the perturbation 
problem~(\ref{GLNA}) has the form
\begin{equation}\label{full}
    \dot{x}= \boldsymbol{N}(x)\boldsymbol{\mu}(x) + \varepsilon\boldsymbol{\tilde{G}}(x,\varepsilon)+ \Omega^{-1/2}B(z)\Gamma,
\end{equation}
where $x=(x_1,x_2,X_1,X_2)^T$, and the critical set, $\widetilde{S}$, that emerges when $\varepsilon = \Omega^{-1}=0$ is
\begin{equation}
    \widetilde{S}:=\{x\in \mathbb{R}^4: \boldsymbol{\mu}(x) =0\}.
\end{equation}
The corresponding projection operator is
\begin{equation}\label{Fullproj}
    \widetilde{\Pi}^{\widetilde{S}_0}:= \boldsymbol{I} - \boldsymbol{N}(\mathcal{D}\boldsymbol{\mu} \cdot \boldsymbol{N})^{-1}\mathcal{D}\boldsymbol{\mu},
\end{equation}
where $\mathcal{D}$ denotes differentiation with respect to $x$ with 
$x:=(x_1,x_2,X_1,X_2)^T$. The projection of the right hand side of (\ref{full}) onto the tangent 
space of $\widetilde{S}_0$ is

\begin{equation}\label{GLNA2}
   \dot{x} = \varepsilon \widetilde{\Pi}^{\widetilde{S}_0} \boldsymbol{\tilde{G}}(x,0)|_{x\in \widetilde{S}_0} + \Omega^{-1/2} \widetilde{\Pi}^{\widetilde{S}_0} B(z)\Gamma|_{x\in \widetilde{S}_0}.
\end{equation}

\begin{remark}
In the nonlinear Langevin regime, the reduced equation may contain a stochastic drift term that is 
$\mathcal{O}(\Omega^{-1})$ (see, \citet{Katz} and \citet{Parsons2017}). Hence, simply projecting onto the tangent 
space of the critical manifold does not yield a sufficient reduction of the Langevin equation unless 
the drift term vanishes or can be ignored. Such a term will also be present in the LNA regime. As 
\citet{Parsons2017} point out, the drift term is not negligible when: the curvature of the slow manifold 
is significant, the curvature effect of the flow field is extreme, or the angle between the fast and 
slow subspace generates a bias in the way a trajectory returns to the slow manifold. It may be possible 
to discard the drift term when $\Omega$ is sufficiently large, but proof of this conjecture is open. 
Hence, (\ref{GLNA2}) holds for systems that have a negligible (or identically zero) drift term. 
\end{remark}

%%%%%
\subsection{The ssLNA for systems in standard form: comparison to previous results}
The reduction method introduced by \citet{Thomas2012} differs from (\ref{GLNA2}). 
For a two-dimensional singularly perturbed problem in the standard form,\footnote{For simplicity, we 
have assumed that $\eps f(x,y)$ contains only terms that are $\mathcal{O}(\eps)$, as this form is 
common in chemical kinetics. The analysis of the more general form can be found in \citet{Wechselberger2020}.} 
\begin{subequations}\label{standard}
\begin{align}
    \dot{x} &= \varepsilon f(x,y),\\
    \dot{y} & =g_0(x,y)+\eps g_1(x,y,\eps),
\end{align}
\end{subequations}
 the critical manifold $S_0:=\{(x,y)\in \mathbb{R}^2:g_0(x,y)=0\}$ attracts nearby trajectories if 
 $g_{0y} <0\;\;\forall (x,y)\in S_0$. Moreover, by the Implicit Function Theorem,  
 $g_{0y} \neq 0\;\;\forall (x,y)\in S_0$ implies the critical manifold is expressible as $y=h(x)$:
 \begin{equation*}
     g_0(x,h(x))=0.
 \end{equation*}
\citet{Thomas2012} construct the ssLNA directly from the non-singular Jacobian\footnote{The notation 
$f_y$ denotes $\partial_y f(x,y)$.} (that corresponds to $0 <\eps$),
\begin{equation*}
   J = \begin{pmatrix}f_x & f_y \\ g_x & g_y \end{pmatrix}, \quad \mathcal{S}=\begin{pmatrix}\mathcal{S}_{slow} \\ \mathcal{S}_{fast}\end{pmatrix}
\end{equation*}
with the a priori requirement that the system be in standard form. From this, they define the maps:
\begin{subequations}\label{MAP1}
\begin{align}
\bar{J} &:= f_x-\cfrac{g_{x}}{g_y}\cdot f_{y}\\
    \bar{\mathcal{S}}_{slow} &:= \mathcal{S}_{slow} - \cfrac{f_y}{g_y}\cdot \mathcal{S}_{fast} = \begin{pmatrix}1 & -f_y/g_y \\ 0 & 0\end{pmatrix}\begin{pmatrix}\mathcal{S}_{slow} \\ \mathcal{S}_{fast}\end{pmatrix}
\end{align}
\end{subequations}
Let $X$ and $Y$ denote the respective fluctuations from the $x$ and $y$. The ssLNA of \citet{Thomas2011} is
\begin{subequations}
\begin{align}
    \dot{x} &= f(x,h(x)),\\
    \dot{X} &= \bar{J}X +(\bar{\mathcal{S}}_{slow}\sqrt{F}\cdot\Gamma)|_{y=h(x)}.
\end{align}
\end{subequations}

In contrast, to derive the deterministic sQSSA from GSPT, we begin with the \textit{singular} Jacobian 
of the layer problem associated with (\ref{standard}). The corresponding projection operator $\Pi^{S_0}$ is
\begin{equation}
    \Pi^{S_0}:=\begin{pmatrix}1 & 0 \\ -g_{0x}/g_{0y}&0\end{pmatrix},
\end{equation}
and again the level set $g_0(x,y)=0$ defines the critical manifold, $S_0$. The perturbation term, $\eps G(z,\eps)$ is
\begin{equation*}
   \eps G(z,\eps) :=\eps \begin{pmatrix}
     f(x,y)\\
     g_1(x,y,\eps)
    \end{pmatrix}
\end{equation*}
and therefore the reduced flow for the mean field is
\begin{equation*}
   \dot{x} =\begin{pmatrix}1 & 0 \\ -g_{0x}/g_{0y}&0\end{pmatrix} \begin{pmatrix}
     f(x,y)\\
     g_1(x,y)
    \end{pmatrix} = \begin{pmatrix}f(x,y)\\ -g_{0x}/g_{0y}\cdot f(x,y)\end{pmatrix}.
\end{equation*}
Again, $g_{0y} \neq 0 \;\;\forall z \in S_0$ implies $y=h(x)$ such that $g_0(x,h(x))=0$. Thus, the sQSSA for $x$ is:
\begin{equation}
    \dot{x} = f(x,h(x)).
\end{equation}

For the corresponding ssLNA, and for two-dimensional systems of the standard form (\ref{standard}), we have
\begin{equation*}
    \boldsymbol{N}(x)=\begin{pmatrix}0 & 0 \\
    1 & 0 \\
    0 & 0 \\ 
    0 & 1\end{pmatrix}, \quad \boldsymbol{\mu}(x) = \begin{pmatrix}
    g_0(x,y)\\
    g_{0x} X + g_{0y} Y
\end{pmatrix},
\end{equation*}
and thus the critical manifold is
\begin{equation*}
    \widetilde{S}:=\{(x,y,X,Y)\in \mathbb{R}^4: g(x,y)=0,\;g_{0x}X+g_{0y}Y=0\}.
\end{equation*}
The perturbation term, $G(z,\eps)$, is
\begin{equation*}
    G(z,\eps):=\begin{pmatrix}f(x,y)\\ g_1(x,y,\eps) \\ f_{x}(x,y)X+f_y(x,y)Y\\ g_{1x}(x,y,\eps) X+g_{1y}(x,y,\eps)Y\end{pmatrix}.
\end{equation*}
Computing $\widetilde{\Pi}^{\widetilde{S}_0}$ from $\boldsymbol{\mu}(x)$ and $\boldsymbol{N}(x)$ and projecting 
$G(z,0)$ onto the tangent space of the critical manifold yields
\begin{subequations}\label{red}
\begin{align}
    \dot{x} &= f(x,h(x))\\
    \dot{X} &= \bigg(f_{x}-f_{y}\cdot\cfrac{g_{0x}}{g_{0y}}\bigg)X +(\Omega^{-1/2}\mathcal{S}_{slow}\sqrt{F}\cdot\Gamma)|_{y=h(x)},\\
    &= \cfrac{\text{d}}{\text{d}x}\bigg(f(x,h(x))\bigg)X+ (\Omega^{-1/2}\mathcal{S}_{slow}\sqrt{F}\cdot \Gamma)|_{y=h(x)}.
\end{align}
\end{subequations}

\begin{remark}
We note that for systems in standard form, an equivalent reduction to (\ref{red}) is given in \citet{Herath} 
and extends to non-autonomous systems.
\end{remark}

Note the difference from the ssLNA of \citet{Thomas2012}. First, we do not map
\begin{equation*}
\mathcal{S}_{slow} \mapsto \mathcal{S}_{slow}-\cfrac{f_y}{g_y}\cdot \mathcal{S}_{fast}.
\end{equation*}
This is a consequence of the fact that our derivation from GSPT begins with the singular Jacobian, which is 
consistent with singular perturbation theory. In contrast, \citet{Thomas2012} began with the perturbed, 
non-singular Jacobian. Consequently, when derived from GSPT, the ssLNA 
contains fewer diffusion terms than the ssLNA of \citet{Thomas2012}. However, for planar systems 
$\mathcal{S}_{slow}$ and $\bar{\mathcal{S}}_{slow}$ should be close whenever $|f_y/g_y| \ll 1$. Hence, 
the difference between the ssLNA of \citet{Thomas2012} and (\ref{red}) should be small when the perturbation 
is in standard form. We note that $\mathcal{S}_{slow}$ is also invariant in the ssLNAs of \citet{Herath} 
and \citet{PAHLAJANI201196}.

Second, observe that
\begin{equation*}
\bar{J} = f_x-\cfrac{g_{x}}{g_y}\cdot f_{y} \neq f_{x}-\cfrac{g_{0x}}{g_{0y}}\cdot f_{y} 
\end{equation*}
unless $g(x,y)$ does not depend on $\eps$, which is not always the case in applications. This difference 
follows from the utilization of the singular Jacobian in derivation. 

%%%%%
\subsection{Benchmark example: The MM reaction mechanism in the limit of small $k_2$ in $(c,p)$  coordinates}
To demonstrate the projection operator methodology on a problem that is in standard form, we analyze 
the MM reaction mechanism in $(p,c)$ coordinates and consider the limit of small 
$k_2$: $k_2 \mapsto \varepsilon \widehat{k}_2.$ In $(p,c)$-coordinates, the deterministic rate equations 
are given by\footnote{In (\ref{PRE}), $s_T$ denotes the total substrate.}
\begin{equation}\label{PRE}
    \begin{pmatrix}
    \dot{p}\\
    \dot{c}
    \end{pmatrix} = N(z)\mu(z) + \varepsilon G(z) :=\begin{pmatrix}0\\ 1\end{pmatrix}(k_1(e_T-c)(s_T-c-p)-k_{-1}c) + \varepsilon \begin{pmatrix}\;\;\;\widehat{k}_2c\\-\widehat{k}_2c\end{pmatrix},
\end{equation}
which is in the standard form (\ref{standard}); $p$ is the slow variable and $c$ is the fast variable. 
The projection matrix is\footnote{Again, $\mu_c$ denotes $\partial_c \mu(p,c)$ and $\mu_p$ denotes
$\partial_p \mu(p,c)$.}
\begin{equation}
    \Pi^{S_0} := \begin{pmatrix}1 & 0 \\ -( \mu_c)^{-1}\mu_p & 0\end{pmatrix},
\end{equation}
and the critical manifold
\begin{equation*}
    S_0 :=\{(p,c)\in \mathbb{R}^2 : \mu(p,c)=0, \;\;0 \leq c \leq e_T,\;\;0 \leq p \leq s_T\}
\end{equation*}
is normally hyperbolic and attracting since
\begin{equation}\label{CM3}
   \langle D\mu,N\rangle = \mu_c =\cfrac{\partial}{\partial c}\; [k_1(e_T-c)(s_T-c-p)-k_{-1}c)] < 0, \quad \forall (c,p)\in S_0.
\end{equation}
Since $\mu_c < 0\;\;\forall (c,p)\in S_0$, it follows from the the Implicit Function Theorem that the 
critical manifold is expressible as $c=y(p)$,
\begin{equation}\label{tQSSA}
    y(p) = \cfrac{k_2}{2}\bigg(s_T+e_T+K_S-p-\sqrt{(s_T+e_T+ K_S-p)^2-4e_T(s_T-p)}\bigg),
\end{equation}
where $K_S=k_{-1}/k_1$. The reduced equation for $p$ is
\begin{equation}
  \dot{p}=k_2y(p).  
\end{equation}
One could also employ the total QSSA (tQSSA) in this case. Again, see \citet{Herath} for excellent 
analysis of the ssLNA in the context of the tQSSA.

We now turn to the reduction of the LNA. The complete perturbation form of the LNA is
\begin{equation}\label{PLNA}
    \begin{pmatrix}
    \dot{p} \\ \dot{c} \\ \dot{X}_p \\ \dot{X}_c
    \end{pmatrix} = \begin{pmatrix}0 & 0 \\
    1 & 0\\
    0 & 0 \\
    0 & 1\end{pmatrix} \begin{pmatrix} \mu(p,c) \\ D\mu(p,c)\cdot X \end{pmatrix} + \varepsilon \widehat{k}_2 \begin{pmatrix}\;\;\;c\\ -c\\ \;\;\;X_c \\-X_c\end{pmatrix} + \Omega^{-1/2} B\cdot \Gamma,
\end{equation}
where $B$ is given by
\begin{equation}
    B:=\begin{pmatrix} 0 & 0 & 0\\ 0 & 0 & 0\\
    0 & 0 &\;\;\;\sqrt{\varepsilon \widehat{k}_2c}\\
    \sqrt{(k_1(e_T-c)(s_T-p-c)}  & -\sqrt{k_{-1}c} & -\sqrt{\varepsilon \widehat{k}_2 c},
    \end{pmatrix}
\end{equation}
and $\mu(p,c)$ in (\ref{PLNA}) is 
\begin{subequations}
\begin{align}
    \mu(p,c) &= k_1(e_T-c)(s_T-c-p)-k_{-1}c,\\
    D\mu(p,c)\cdot X &= \mu_pX_p+\mu_cX_c.
    \end{align}
\end{subequations}
The critical manifold,
\begin{equation}
   \widetilde{S}:=\{(p,c,X_p,X_c) \in \mathbb{R}^4 : \mu(p,c)=0 \;\;\& \;\;\mu_pX_p+\mu_cX_c =0\},
\end{equation}
is normally hyperbolic and attracting.
Proceeding in the usual way by calculating $\widetilde{\Pi}^{\widetilde{S}_0}$, the 
reduced equation for $X_p$ is
\begin{equation}\label{pred}
    \text{d}{X}_p = k_2X_c\text{d}t + \Omega^{-1/2}\sqrt{k_2y(p)}\; \text{d}W_3(t).
\end{equation}
To eliminate $X_c$ from (\ref{pred}) we invoke the critical manifold relationship
\begin{equation}
    X_c = -\mu_c^{-1}\mu_pX_p\;\;\text{with}\;\;c=y(p),
\end{equation}
which yields
\begin{equation}\label{NtQSSA}
    \text{d}{X}_p = k_2 y'(p) X_p\text{d}t + \Omega^{-1/2}\sqrt{k_2y(p)} \;\text{d}W_3(t),
\end{equation}
where ``$y'(p)$" denotes $\cfrac{\text{d}y}{\text{d}p}$. Interestingly, it is worthwhile noting 
that equations (\ref{tQSSA}) and (\ref{NtQSSA}) are equivalent to the tQSSA in the linear noise 
regime.

%%%%%
\subsection{Estimating conditions for the QSS: The MM reaction mechanism with feedback in the 
limit of small $k_2$ and $k_3$}
In this subsection, we analyze the QSS behavior of the MM reaction mechanism with feedback:
\begin{equation}\label{mmF}
    \ce{S + E <=>[$k_1$][$k_{-1}$] C ->[$k_2$] E + P }, \quad \ce{P ->[$k_3$] S}.
\end{equation}
In $(p,c)$-coordinates the reaction is modelled by the ODE system
\begin{subequations}
\begin{align}
\dot{p} &=k_2c-k_3p,\\
\dot{c} &=k_1(e_T-c)(s_T-c-p)-(k_{-1}+k_2)c,
\end{align}
\end{subequations}
which admits a nontrivial steady-state solution at $(p,c)=(p_{SS},c_{SS})$. Furthermore, small 
$k_2$ and $k_3$ defines a singularly perturbed system in the standard form (\ref{standard}):
\begin{subequations}
\begin{align*}
\dot{p} &=\eps \widehat{k}_2c-\eps \widehat{k}_3p,\\
\dot{c} &=k_1(e_T-c)(s_T-c-p)-(k_{-1}+\eps \widehat{k}_2)c.
\end{align*}
\end{subequations}

The LNA approximation includes the randomly fluctuating departure from the mean field (\ref{mmF}),
\begin{equation}
 \begin{pmatrix}
 \dot{X}_p\\\dot{X}_c
 \end{pmatrix}  = J\begin{pmatrix} X_p\\X_c\end{pmatrix} + \Omega^{-1/2}\begin{pmatrix} 0 & 0 & \sqrt{k_2c} & -\sqrt{k_3p}\\ \sqrt{k_1(e_T-c)(s_T-c-p)} & -\sqrt{k_{-1}c} & -\sqrt{k_2c} & 0\end{pmatrix}\cdot \Gamma
\end{equation}
where the Jacobian, $J$, is given by
\begin{equation*}
    J:=\begin{pmatrix} -k_3 & k_2\\ -k_1(e_T-c) & -k_1(s_T-c-p)-k_1(e_T-c)-k_{-1}-k_2\end{pmatrix}.
\end{equation*}
Under QSS conditions, the covariance matrix, $\Sigma$, of the LNA satisfies the Lyapunov equation,
\begin{equation*}
    J\Sigma + \Sigma J^{tr.} = -\Omega^{-1} \mathcal{S}F\mathcal{S}^{tr.}.
\end{equation*}
The variance of the slow variable, $p$, is  $\Sigma(1,1)$.

\begin{figure}[htb!]
    \centering
    \includegraphics[scale=0.75]{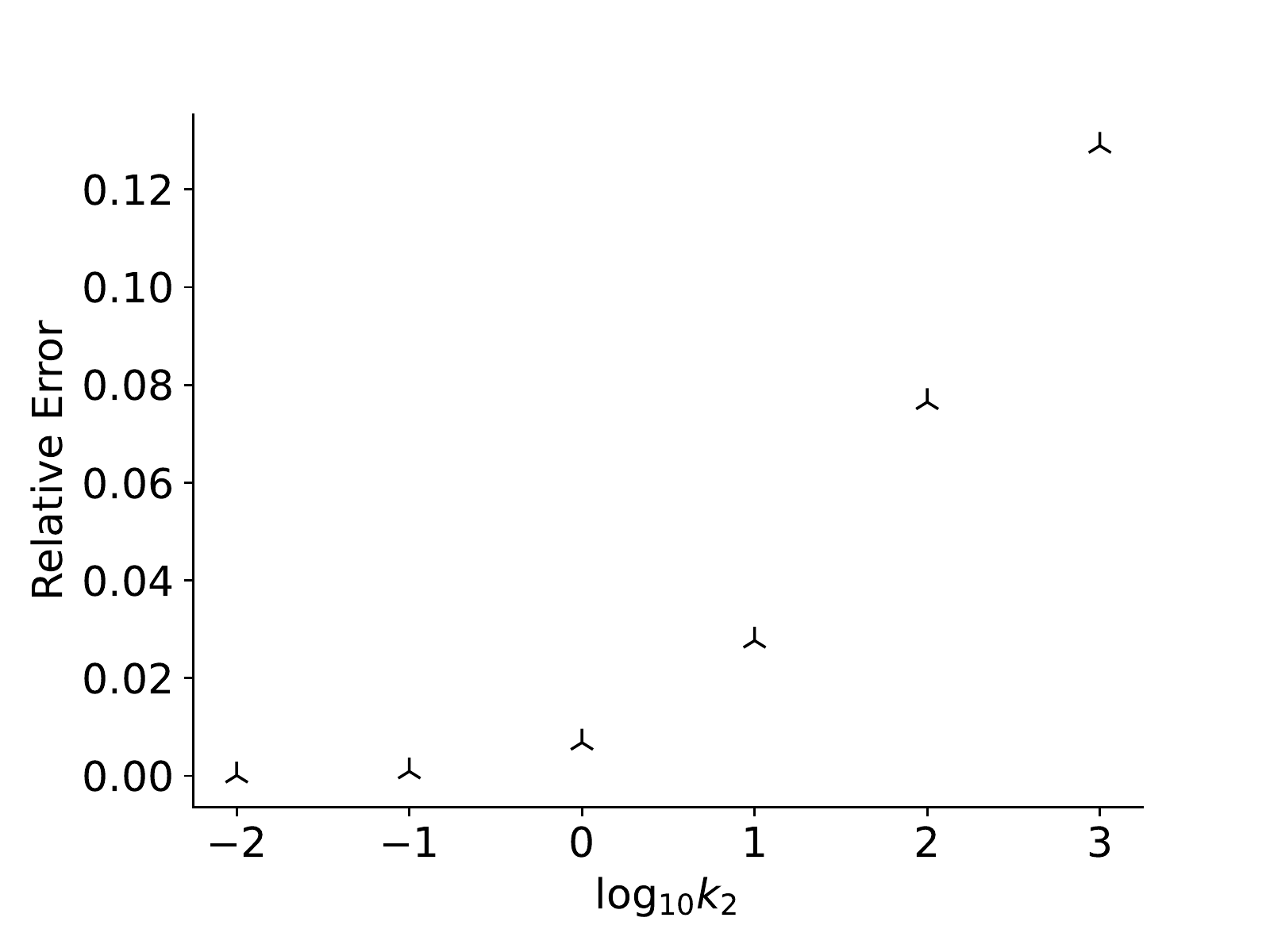}
    \caption{\textbf{The relative error between the QSS variance of GSPT-derived ssLNA and 
    QSS variance of the LNA for $p$ decreases as $k_2,k_3\to 0$}. The $y$-axis is the relative 
    error $|\Sigma(1,1)-\sigma_p|/\Sigma(1,1)$; the $x$-axis is $\log_{10} k_2$. Parameters 
    (in arbitrary units) are: $e_T=1000.0$, $s_T=2000.0$, $k_1=1.0$, $k_{-1}=1.0$ and $\Omega=1.0$ 
    The parameters $k_2$ and $k_3$ are equal range and from $10^{3}$ to $10^{-2}$.
    \label{fig:Rel}}
\end{figure}

The corresponding ssLNA is
\begin{subequations}
\begin{align}
\dot{p}&=k_2y(p) -k_3p\\
\text{d}X_p &= (k_2y'(p) -k_3)X_p\text{d}t +\Omega^{-1/2}\sqrt{k_2y(p)}\;\text{d}W_3 - \Omega^{-1/2}\sqrt{k_3p}\;\text{d}W_4,
\end{align}
\end{subequations}
and under steady-state conditions the variance is
\begin{equation}\label{var}
    \sigma_p:=\cfrac{1}{2}\bigg(\cfrac{k_2y(p)+k_3p}{|k_2y'(p)-k_3|}\bigg)\bigg|_{p=p_{SS}}.
\end{equation}
Numerical simulations confirm that (\ref{var}) is an excellent approximation to $\Sigma(1,1)$ 
as $k_3,k_2 \to 0$ (see, {{\sc Figure}}~\ref{fig:Rel}).

%%%%%%%%%%
\section{Reduction of the CME: Intimations from the linear noise regime}

In this section, we discuss the reduction of the CME for the MM reaction mechanism and its relationship 
to singular perturbations and critical manifolds. Specifically, we address the presence of transcritical 
bifurcations in the linear noise regime and illustrate that knowledge of the critical manifold can 
assist in avoiding erroneous conclusions concerning the validity of the stochastic QSSA. 

%%%%%
\subsection{Dynamic bifurcations and the Segel--Slemrod sQSS condition}
The CME for the MM reaction mechanism is
\begin{multline}\label{mmCME}
    \cfrac{\partial}{\partial t}P(n_S,n_C,t)=\bigg[\cfrac{k_1}{\Omega}(\mathbb{E}_S^{+1}\mathbb{E}_C^{-1}n_S(n_{e_T}-n_C)  \\ + k_{-1}(\mathbb{E}_S^{+1}\mathbb{E}_C^{-1}-1)n_C + k_2(\mathbb{E}_C^{-1}-1)n_C\bigg]P(n_S,n_C,t).
\end{multline}
where $n_{e_T}$ denotes the total number of enzyme molecules, and $P(n_S,n_C,t)$ is the probability 
of finding the system with $n_S$ substrate molecules and $n_C$ complex molecules at time $t$.

The homologous sQSS reduction of (\ref{mmCME}) is as follows. Given that there are $n_S$ substrate 
molecules at time $t$, the probability that one one product molecule forms in an infinitesimal 
window $[t,t+\text{d}t)$ is
\begin{equation}\label{prop}
    P(n_S-1, t+\text{d}t|n_S,t):= a(n_S)\text{d}t = \cfrac{k_2e_Tn_S}{K_M+n_S/\Omega}\;\text{d}t,
\end{equation}
where the propensity function, $a(n_S)$, is adopted from deterministic sQSSA rate law, and the 
reduced CME is
\begin{equation}\label{stochQSSA}
    \cfrac{\partial}{\partial t}P(n_S,t)= (\mathbb{E}_S^{+1}-1)\cfrac{k_2e_Tn_S}{K_M+n_S/\Omega}P(n_S,t).
\end{equation}
In what follows, for simplicity, we set $\Omega = 1$ and work in arbitrary units; however, we perform our simulations with a large number of molecules.

Numerical work by \citet{Sanft} suggests that the Segel--Slemrod condition (expressed in terms of 
stochastic rate constants)
\begin{equation}\label{SSc}
    e_T \ll K_M +s_T
\end{equation}
ensures the validity of the stochastic sQSSA (\ref{stochQSSA}). This is surprising, especially 
since the long-time validity of the deterministic sQSSA for the MM reaction mechanism requires 
$e_T/K_M\ll 1$ \cite{EILERTSEN2020}, which is more restrictive than the Segel and Slemrod condition. 
However, excellent (and extremely rigorous) work by \citet{KangWon2017} disputes this claim. In 
the stochastic regime, \citet{KangWon2017} concluded that $e_T \ll K_M$, which is in agreement 
with the deterministic qualifier. A similar conclusion was drawn by~\citet{Mastny2007}.

Importantly, although the deterministic and stochastic QSS reductions of the MM mechanism are 
justified via singular perturbation theory, the rigorous derivation of the sQSSA from singular 
perturbation was only recently established~\citet{Goeke2012}. This raises the question: given 
what we now understand about the bifurcation structure of the critical set associated with the 
deterministic MM reaction mechanism, what consequence(s) does this have on the stochastic QSS 
reduction? More specifically, does the Segel and Slemrod condition guarantee that the stochastic 
sQSSA will remain accurate for all time, or is the more restrictive condition derived by
\citet{KangWon2017} and~\citet{Mastny2007} necessary to ensure the accuracy of the stochastic
sQSSA?

To answer this question, we note that the perturbation problem corresponding to small $k_{-1}$ 
and $k_2$ is of the form (\ref{standard}):
\begin{subequations}
\begin{align}
    \dot{p} &= \eps \widehat{k}_2 c\\
    \dot{c} &= k_1(e_T-c)(s_T-c-p) - \eps(\widehat{k}_{-1}+\widehat{k}_{2}).
\end{align}
\end{subequations}
If $e_T \ll s_T=s(0) $, then the Fenichel reduction is formulated by projecting the perturbation 
onto the tangent space of $\mathcal{S}_a^r$, $T_z\mathcal{S}_a^r$:
\begin{subequations}
\begin{align}
    \dot{p} &= k_2e_T,\\
    \dot{c} &= 0.
\end{align}
\end{subequations}
This approximation does not hold for all-time: eventually the trajectory follows $\mathcal{S}_a^b$, 
and the Fenichel reduction is
\begin{subequations}
\begin{align}
\dot{p} &= k_2(s_T-p),\\
\dot{c} &= -k_2(s_T-p).
\end{align}
\end{subequations}
The behavior of the reduction in small neighborhoods containing the bifurcation point is beyond 
the scope of this paper. In general, one must defer to non-classical methods to derive scaling 
laws near the bifurcation point (see, \citet{Berglund,krupa2001}).

As $\Omega$ shrinks and fluctuations emerge, the LNA holds sway. The presence of a bifurcation
point in the critical set is not too restrictive in this case. The ssLNA obtained via projection 
onto $\mathcal{S}_a^r$ is
\begin{subequations}\label{ss1}
\begin{align}
    \dot{p} &= k_2e_T,\\
    \text{d}X_p &= \Omega^{-1/2}\sqrt{k_2e_T}\;\text{d}W_3.
\end{align}
\end{subequations}
Note the relationship between the mean and variance. Likewise, projecting onto $T_z\mathcal{S}_a^b$ yields
\begin{subequations}\label{ss2}
\begin{align}
    \dot{p} &= k_2(s_T-p),\\
    \text{d}X_p &= -k_2 X_p \text{d}t + \Omega^{-1/2}\sqrt{k_2(s_T-p)}\;\text{d}W_3.
\end{align}
\end{subequations}

As $\Omega \to 0$ the CME prevails as the physically relevant model. The ssLNA (\ref{ss1}) is a 
Gaussian process with equal mean and variance. In the CME regime, the reaction mechanism on 
$\mathcal{S}_a^r$ is equivalent to
\begin{align}\label{redCHEM}
    \emptyset \ce{->[$\lambda$] P},
\end{align}
where $\lambda=k_2n_{e_T}$ and $n_{e_T}$ denotes the total number (bound or unbound) of enzyme 
molecules. The CME that describes (\ref{redCHEM}) is solvable. Let $P(N,t)$ denote the probability 
that there are $N$ product molecules at time $t$. Then,
\begin{equation}\label{redP}
    P(N,t) = \exp{(-\lambda t)}\cdot \cfrac{(\lambda t)^N}{N!}.
\end{equation}
Note the consistency with (\ref{ss1}). The Poisson jump process is approximately Gaussian when 
the system size is sufficiently large.

Unfortunately, (\ref{redP}) is not valid for all-time, and it is necessary to ascertain the 
range of its validity. More precisely, we ask: How long (on average) from the onset of the 
reaction does it take before (\ref{redP2}) is valid? Since (\ref{redCHEM}) is a Poisson process 
the jump times, $t_N$, are gamma-distributed:
\begin{equation*}
    t_N \sim \lambda^N \exp{(-\lambda t)} \cdot \cfrac{t^{N-1}}{(N-1)!}.
\end{equation*}
Let $n_{s_T}$ denote the total number of substrate molecules. The average time it takes to 
produce $N^*=n_{S_T}-n_{e_T}$ product molecules is
\begin{equation*}
   \langle t_{N^*}\rangle = \cfrac{N^*}{\lambda},
\end{equation*}
which is exactly homologous to the deterministic scenario.

Moving on, once $N = n_{S_T}-n_{e_T}$ we arrive at the intersection of the critical branches, 
$\mathcal{S}_a^r \cap \mathcal{S}_a^b$. At this point, no substrate molecules remain and the 
formation product is synonymous with the depletion of $c$:
\begin{align}\label{redCHEM2}
    C \ce{->[$k_2$] P}.
\end{align}
Once again, the CME associated with (\ref{redCHEM2}) is solvable:
\begin{equation}\label{redP2}
P(n_{S_T}-n_c,t) = P(n_c,t)= \exp(-k_2n_c t)\begin{pmatrix}n_{e_T} \\ n_c\end{pmatrix}(1-\exp(-k_2t))^{(n_{e_T}-n_c)},
\end{equation}
where $n_c$ denotes the number of complex molecules, and time has been translated so that:
\begin{equation*}
    P(n_{S_T}-n_{e_T},0) =1.
\end{equation*}

\begin{figure}[htb!]
    \centering
    \includegraphics[scale=0.5]{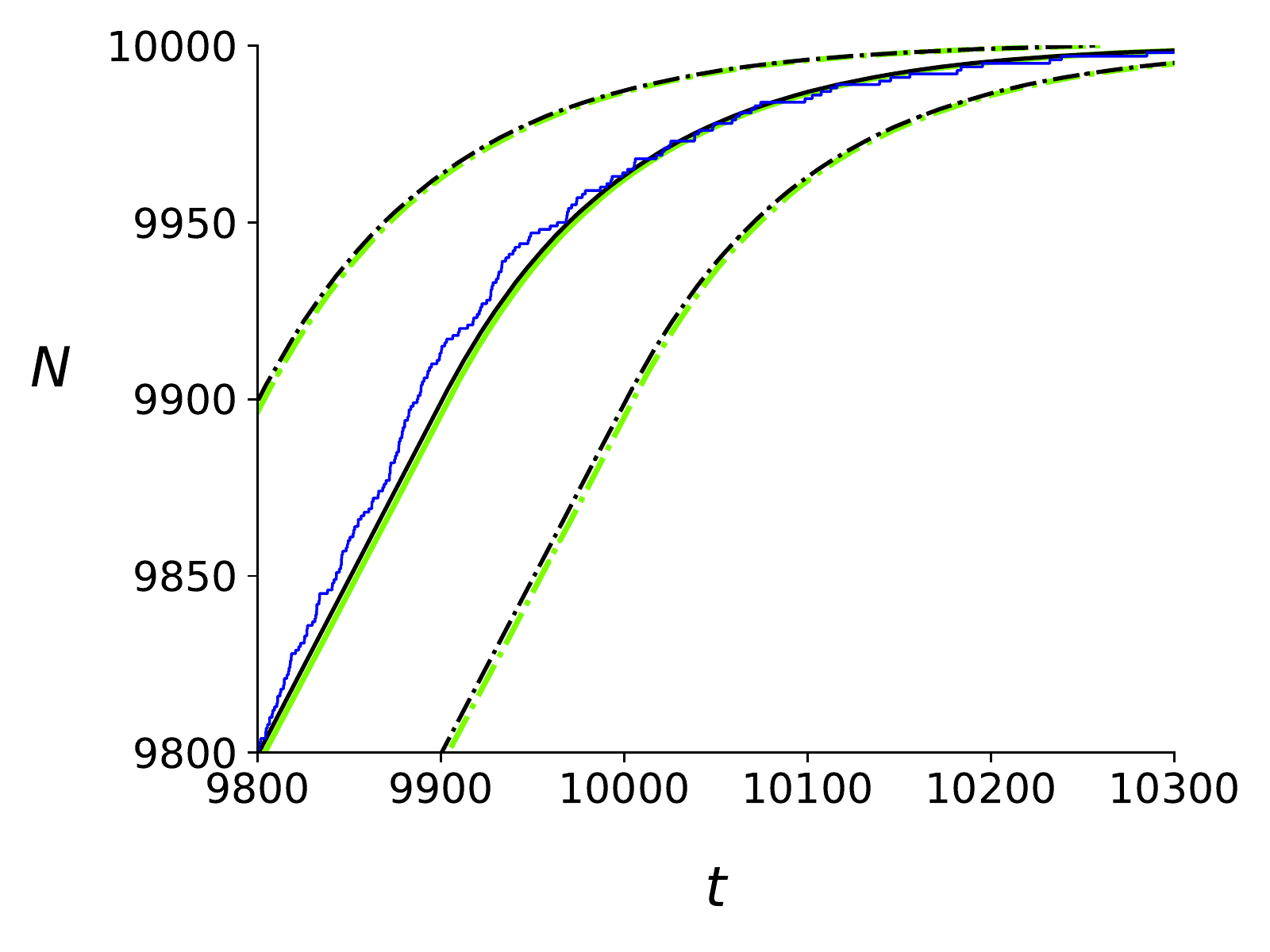}
    \includegraphics[scale=0.5]{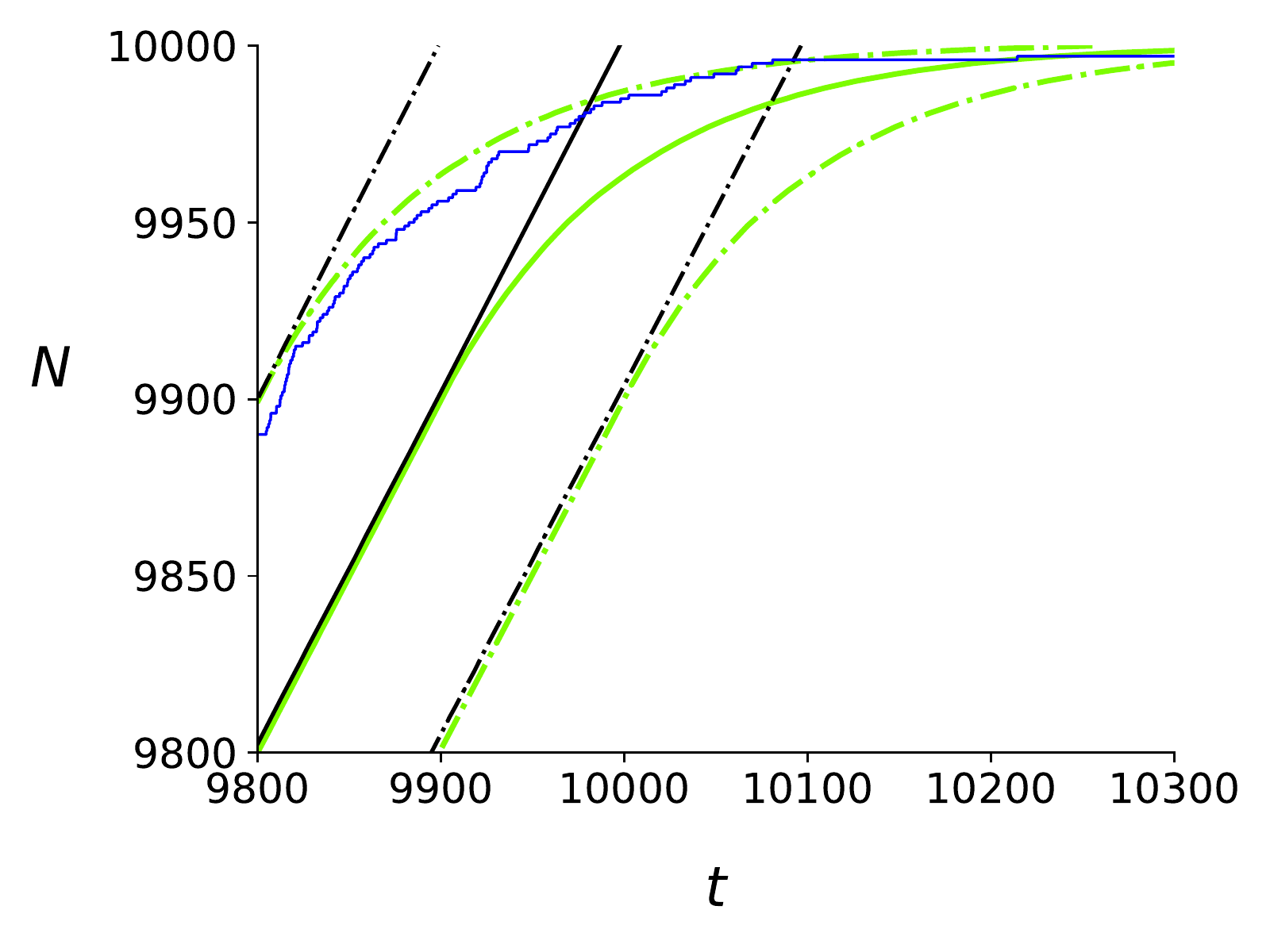}
    \caption{\textbf{The stochastic sQSSA (\ref{prop}) will fail near the bifurcation point if $e_T/K_M \gg 1$.} In both 
    panels, the solid green curve is the numerically-computed mean of the timecourse for $N$ (the number of product molecules) 
    obtained from $1000$ simulations generated by the Gillespie algorithm; the dashed/dotted green curves demarcate the mean 
    $\pm$ the standard deviation. The solid black curve is the numerically-computed mean of the time course for $N$ (the number 
    of product molecules) obtained from $1000$ simulations generated by the Gillespie algorithm equipped with a QSS-derived 
    propensity function; the blue line shows one such randomly picked simulation; the dashed/dotted black lines demarcate 
    the mean $\pm$ the standard deviation. In both simulations, the parameters (in arbitrary units) are: $n_{s_T}=10000$, 
    $n_{e_T}=100$, $k_1=1000.0$, $k_{-1}=0.01$, $k_2=0.01$ and $\Omega = 1.0$. $n_{s}(0)=n_{s_T}$, $n_c(0)=0$, $n_e(0)=n_{e_T}$, 
    and $N(0)=0$. Note that $e_T/(K_M+s_T)\approx 0.01$ and therefore the Segel-Slemrod (\ref{SSc}) condition holds.  
    {\sc Left} panel: The black solid and dashed/dotted lines are obtained from Gillespie algorithm equipped with the 
    propensity function (\ref{broken}); the first two statistical moments are practically indistinguishable. 
    {\sc Right} panel: The black solid and dashed/dotted lines are obtained from Gillespie algorithm equipped with the 
    propensity function (\ref{prop}), and the stochastic sQSSA fails near $N=N^*$.
    \label{fig:STOCH}}
\end{figure}

The question that remains is: How should the Gillespie algorithm be modified to reduce the 
computational complexity when $k_2$ and $k_{-1}$ are sufficiently small? The above analysis 
indicates that $P(N,t)$ depends on the number of product molecules present at a given time 
in the reaction. Specifically, $P(N,t)$ depends on whether or not $N <N^*$. To modify the 
Gillespie algorithm, observe that the propensity function $a(N)$ for product formation -- at 
any given time -- depends on the number of product molecules, $N$, present at time $t$. Thus,
\begin{equation}\label{broken}
   a(N)=  \begin{cases}
           k_2n_{e_T}, \qquad \text{if}\;\; N < N^*,\\
           k_2n_c, \qquad \;\;\text{if}\;\; N\geq N^*.
    \end{cases}
\end{equation}
Numerical simulations support the results of our analysis, and demonstrate that the 
Segel and Slemrod condition does not imply the validity of the stochastic sQSSA 
(see {{\sc Figure}}~\ref{fig:STOCH}). 

We note that one can employ the reduction technique of \citet{Thomas2012}. In general, the 
ssLNA of \citet{Thomas2012} will be close (in the asymptotic sense) to (\ref{ss1})--(\ref{ss2}), 
but will be more complicated due to the presence of additional diffusion terms. The simplicity 
of the GSPT-derived ssLNA (\ref{ss1})--(\ref{ss2}) helps to explain the insufficiency of 
the Segel--Slemrod condition for the validity of the stochastic sQSSA in the linear noise 
regime, thereby validating the results of~\citet{KangWon2017} and~\citet{Mastny2007} 
from the context of GSPT.

As a final remark, we mention that the bifurcation point can also be handled with appropriate utilization of the tQSSA. 
Although the treatment of bifurcation points has so far not been addressed in the stochastic tQSSA literature, several 
rigorous studies suggest that the stochastic tQSSA is superior to the sQSSA in the CME and LNA regimes. Rigorous analyses 
of the stochastic tQSSA are found in~\cite{kim2014,kim2020,kim2015,MacNamara2008,Tyson2008}.

%%%%%%%%%%
\section{Discussion}\label{sec5}
The primary contribution of this work is the derivation of the ssLNA in a way that is consistent with 
Fenichel theory \cite{Fenichel1979}, that is, the projection of a perturbation term onto the tangent 
space of a normally hyperbolic critical manifold. Our derivation explains the origin of the differences 
between the the ssLNAs reported in \citet{Thomas2012}, \citet{PAHLAJANI201196}, and \citet{Herath}.

By re-deriving the ssLNA directly from GSPT, we illustrated how the GSPT-derived ssLNA can be extended 
to singular perturbation problems where normal hyperbolicity fails and classical Fenichel theory breaks 
down. To the best of our knowledge, this is the first extension of the ssLNA to singular perturbation 
problems that contain a transcritical bifurcation. 

Finally, let us remark on the possible special role of the standard form in the reduction of the CME. 
In their derivation of the ssLNA, \citet{Thomas2012} shared the following insight: the mapping
\begin{equation*}
\mathcal{S}_{slow} \mapsto \mathcal{S}_{slow}-\cfrac{f_y}{g_y}\cdot \mathcal{S}_{fast}
\end{equation*}
does not result in physically meaningful slow variable stoichiometry in the CME regime. However, as we 
pointed out, when the system is truly in standard form (again, the MM reaction mechanism with small 
$e_T$ does not technically qualify), the stoichiometry component of the slow variable, $\mathcal{S}_{slow}$ 
is invariant: $\mathcal{S}_{slow}\mapsto \mathcal{S}_{slow}$. This suggests that singularly perturbed 
systems in standard form\footnote{In addition, the drift term $\nu \mathcal{D}(x)$ introduced by~\citet{Katz} 
will presumably be zero when the system is in standard form.} might, in some way, be amenable to 
QSS reduction in the CME regime. However, this hypothesis warrants further investigation.

%%%%%%%%%%
\section*{Acknowledgements}
We are grateful to Dr. Ramon Grima (University of Edinburgh), for providing critical
comments in an early draft of this manuscript. JE was supported by the University of 
Michigan Postdoctoral Pediatric Endocrinology and Diabetes Training Program 
``Developmental Origins of Metabolic Disorder'' (NIH/NIDDK Grant: T32 DK071212). 

%%%%%%%%%%
%\section*{Data Availability}
%All new data created or analyzed in this study is available throughout the paper. 
%Numerical recipes to run simulations shown in {\sc Figure 2} are openly available 
%on \url{https://github.com/santiago-schnell/ss-Linear-Noise-Approximation}.

%% The Appendices part is started with the command \appendix;
%% appendix sections are then done as normal sections
%% \appendix

%% \section{}
%% \label{}

%% References
%%
%% Following citation commands can be used in the body text:
%% Usage of \cite is as follows:
%%   \cite{key}          ==>>  [#]
%%   \cite[chap. 2]{key} ==>>  [#, chap. 2]
%%   \citet{key}         ==>>  Author [#]

%% References with bibTeX database:

 %\bibliographystyle{model1-num-names}

%% New version of the num-names style

%\bibliographystyle{elsarticle-num-names}
\%bibliography{biblio.bib}

%% Authors are advised to submit their bibtex database files. They are
%% requested to list a bibtex style file in the manuscript if they do
%% not want to use model1-num-names.bst.

%% References without bibTeX database:

\end{document}